\documentclass[aps,prl,twocolumn,superscriptaddress,showpacs,floatfix]{revtex4-1}
\usepackage{graphicx,epsfig,amsmath}

\begin{document}


\title{Nonsaturating Dephasing Time at Low Temperature in an Open Quantum Dot}


\author{I. G. Rau}
	\altaffiliation{Present address: IBM Research, San Jose, CA 95120}
	\affiliation{Department of Applied Physics, Stanford University, Stanford, California 94305, USA}

\author{S. Amasha}
	\altaffiliation{Present address: MIT Lincoln Laboratory, Lexington, MA 02420}
	\affiliation{Department of Physics, Stanford University, Stanford, California 94305, USA}

\author{M. Grobis}
	\altaffiliation{Present address: HGST, San Jose, CA 95135}
	\affiliation{Department of Physics, Stanford University, Stanford, California 94305, USA}	

\author{R. M. Potok}
	\altaffiliation{Present address:  Solum Inc., Mountain View, CA 94043}
	\affiliation{Department of Physics, Stanford University, Stanford, California 94305, USA}
	\affiliation{Department of Physics, Harvard University, Cambridge, Massachusetts 02138, USA}

\author{Hadas Shtrikman} 
	\affiliation{Department of Condensed Matter Physics, Weizmann Institute of Science, Rehovot 96100, Israel}
	
\author{D. Goldhaber-Gordon} 
	\email{goldhaber-gordon@stanford.edu}
	\affiliation{Department of Physics, Stanford University, Stanford, California 94305, USA}


\begin{abstract}
    
       We report measurements of the electron dephasing time extracted from the weak localization (WL) correction to the average conductance in an open AlGaAs/GaAs quantum dot from 1 K to 13 mK.  In agreement with theoretical predictions but in contrast with previous measurements in quantum dots, the extracted dephasing time does not saturate at the lowest temperatures. We find that the dephasing time follows an inverse linear power law with temperature. We determine that the extraction of the dephasing time from WL is applicable down to our lowest temperatures, but extraction from finite magnetic field conductance fluctuations is complicated by charging effects below $13$ mK.
     
\end{abstract}

\pacs{73.23.-b, 03.65.Yz, 73.20.Fz}  

\maketitle

Understanding electron dephasing in mesoscopic systems allows us to quantify the effect of  interactions between a quantum system and its environment. This is a necessary step in answering fundamental questions about the mechanisms responsible for the destruction of quantum mechanical coherence and for turning the quantum behavior of microscopic systems into the classical behavior of macroscopic systems. Conversely, understanding dephasing is crucial for developing techniques to accurately manipulate coherent quantum states for the purposes of quantum computation. 

The time scale over which electrons maintain their quantum behavior is the dephasing time $\tau_{\phi}$. Theories (e.g. \cite{Altshuler1985, Zheng1996}) of the different microscopic mechanisms for dephasing of electrons in solids, such as interactions with phonons, other electrons, or magnetic impurities, predict the same low temperature behavior:  a power-law divergence of the dephasing time, the details of which depend on factors such as sample dimensionality and disorder. Although the dephasing time should theoretically diverge in the limit $T \rightarrow 0$ \cite{Altshuler1982, Stern1990}, in experiments in quantum dots it appears to saturate \cite{Clarke1995, Bird1995, Huibers1999, Micolich2001, Hackens2005}.

 Experimentally, the dephasing time can be extracted from electrical transport measurement via Aharonov-Bohm oscillations, universal conductance fluctuations (UCFs), or weak localization (WL): all manifestations of quantum interference. Dephasing times have been extracted in 0-, 1- and 2-dimensional samples with a variety of couplings to the environment, in several different materials  \cite{Eshkol2006,Mohanty1997,Anthore2003,Pierre2003,Clarke1995, Bird1995, Huibers1999, Micolich2001, Hackens2005}. While the results in 2D samples \cite{Eshkol2006} 
follow the expected theoretical prediction \cite{Narozhny2002}, the situation for 1D and 0D systems is considerably more complicated.

Early experiments in disordered metal wires used WL to extract a dephasing time that saturated at low temperatures 
\cite{Mohanty1997}. This prompted the suggestion that zero point fluctuations could be a mechanism for dephasing \cite{Golubev1998, Cedraschi2000}. The physical validity of this explanation was highly debated \cite{Aleiner1998, Aleiner1999}. Theoretical work \cite{Goppert2002, Zarand2004} as well as more recent experiments indicated a link between magnetic impurity scattering of electrons and the saturation of the extracted dephasing time \cite{Anthore2003}.  In the absence of magnetic impurities, a continued increase of the dephasing time down to the lowest accessible temperatures was observed \cite{Pierre2003}, in agreement with the expected behavior for dephasing caused by e-e interaction.

Measurements of 0D systems such as semiconductor quantum dots in GaAs and InGaAs have consistently extracted a dephasing time that saturated below $50-100$ mK \cite{Clarke1995, Bird1995, Huibers1999, Micolich2001, Hackens2005}. In some cases this was attributed to the failure of the semiclassical model used \cite{Clarke1995}  and to the discreteness of the dot spectrum \cite{Bird1995, Micolich2001} when $kT\leq \Delta$. When $\tau_{\phi}$ was extracted from the finite magnetic field variance of the UCFs, it was seen to saturate at a value $\tau_{\phi}^{\rm{SAT}}\approx \tau_{\rm{D}}$ (the dwell time of the electrons on the quantum dot) \cite{Hackens2005}. Hackens \textit{et al.} proposed an empirical formula connecting the temperature $T_{SAT}$ at which the onset of the saturation occurs  to the single particle level spacing $\Delta$ of their measured quantum dots.  They also noted that same correlation between $\tau_{\phi}^{\rm{SAT}}$ and $ \tau_{\rm{D}}$ applied to the measurement results of Ref. \cite{Huibers1999}, although the method used in that experiment ($\tau_{\phi}$ is extracted from the average conductance at zero magnetic field) explicitly accounts for the contribution of the finite time electrons spend on the dot.

In this letter we examine the low temperature behavior of the dephasing time in a $2.6~\mu\mbox{m}^2$  quantum dot with single-mode quantum point contact (QPC) leads down to $13$ mK, 2-3 times lower than previous experiments. We find that the temperature dependence follows a $T^{-1}$ law with a small contribution from $T^{-2}$, similar to the results of \cite{Huibers1998, Hackens2005}. However, unlike all previous measurements, we observe that the dephasing time increases monotonically with decreasing temperature over the entire temperature range.  Our lowest experimentally accessible temperature of $13$ mK is well below the value for onset of the saturation observed in the experiments in \cite{Huibers1999}, as well as the  value $T_{\rm{SAT}}=118$ mK suggested by the empirical formula of Ref.~\cite{Hackens2005} applied to our dot.

\begin{figure}
\begin{center}
\includegraphics[width=8.0cm, keepaspectratio=true]{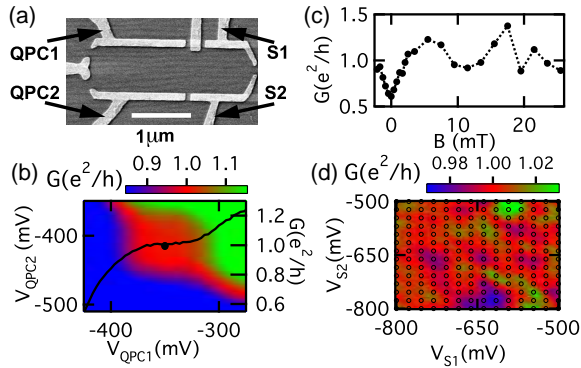}
\end{center}

\caption{(color online) (a) Electron micrograph of the quantum dot indicating the gates used to tune the quantum point contacts (QPC1 and QPC2) and the shape of the dot (S1 and S2). (b) The $1~e^2/h$ conductance plateau at $T= 960$ ~mK and $B= 22$~mT. The black marker indicates the center of the plateau and the black trace (right axis) is a horizontal cut through the 2D conductance plot for the $V_{\rm{QPC2}}$ value corresponding to the black marker.  (c) Dot conductance as a function of magnetic field for $G_{\rm{QPC1,2}}=2~e^2/h$ shows UCFs around  $1~e^2/h$. The dashed line is a guide to the eyes (d) Dot conductance at $720$~mK and $22$~mT as a function of $V_{\rm{S1}}$ and $V_{\rm{S2}}$. The black circles indicate the 196 points used for averaging. }
\label{fig:fig1}
\end{figure}

The quantum dot is fabricated on an AlGaAs/GaAs heterostructure with a two dimensional electron gas  ($n=2\times10^{11}~\mbox{cm}^{-2}$, $\mu=2\times10^6~\mbox{cm}^2/\mbox{Vs}$) situated 68 nm below the surface. Negative voltages applied to the gate electrodes define the quantum dot. The gates labeled QPC1 and QPC2 in Fig.~\ref{fig:fig1}(a) tune the coupling of the electrons in the quantum dot  to the electrons in the extended 2DEG regions that serve as leads. The gates labeled $S1$ and $S2$  affect the shape and the area of the quantum dot  \cite{Huibers1998, Huibers19982}. The measurements are performed in a dilution refrigerator using standard lock-in techniques. The electron temperature is determined using Coulomb Blockade thermometry in an adjacent small ($\approx$20 electron) quantum dot. The temperature range is $13$ mK to $1$ K and the bias across the device is kept below $kT/e$ for all temperatures. 
Because the device size $L= 2.5 \mu$m is shorter than the elastic scattering length of the 2DEG  $l_{\rm{elastic}}=15~\mu$m,  we conclude that transport through the dot is ballistic. Furthermore, the ergodic time $\tau_{\rm{erg}}=\frac{L}{v_{\rm{F}}}=8 $ps is shorter than the thermal time $\tau_{\rm{thermal}}=\frac{\hbar}{k_B T}$ over the whole temperature range (equivalently the Thouless energy $\frac{\hbar}{\tau_{\rm{erg}}} > k_BT$), from which we conclude that the quantum dot is 0D \cite{beenakker.vanhouten}.

We use a method based on the non-interacting random matrix theory (RMT) results for the distributed $\sigma$-lead model \cite{Brouwer1997} to extract the dephasing time from the experimentally measured conductance G  of the quantum dot with one fully transmitting spin-degenerate mode in each QPC ($N=1$  and $G_{\rm{QPC1,2}}=2e^2/h$).  For quantum dots with many transmitting modes ($N\gg 1$) in the QPCs, RMT predicts statistical properties of conductance such as its average and its variance. Although there is no explicit theory for quantum dots with $N=1$, previous experiments in this regime have shown that the quantum dot conductance distributions are well described by RMT \cite{Huibers19982}. To connect our measurements to theory we use gate voltages $V_{S1}$ and $V_{S2}$ to vary the shape of the dot and thus collect conductance measurements from an ensemble of dot shapes. We then use both the ensemble-averaged conductance and the variance to extract the dephasing time in this system. 

To identify the gate voltage range for which the two QPCs are open to a single spin-degenerate mode we sweep $V_{\rm{QPC1}}$ and $V_{\rm{QPC2}}$  at a magnetic field ($20$ mT) sufficient to break time reversal symmetry and at a temperature ($960$ mK) where the UCFs are suppressed by thermal averaging and do not obscure the flat region of the plateau.  Fig. \ref{fig:fig1}(b) shows such a sweep where we can identify a plateau at $G=1e^2/h$, corresponding to the ohmic series conductance of two QPCs, each with conductance $2e^2/h$. The middle of the plateau is indicated by the black dot in Fig. \ref{fig:fig1}(c) and is referred to as QPC setting B. To check the robustness of our results against variations in the QPC settings, we choose two other points (QPC settings A and C) around the middle of the plateau as explained in detail in~\cite{SI}. For these three settings, the reflection coefficients of the two QPCs are  each $~1-2\%$ \cite{SI}. As a function of magnetic field, with the two QPCs tuned to the middle of this plateau, $G$ shows conductance fluctuations about an average of $e^2/h$, as illustrated for a specific shape of the dot in Fig.~\ref{fig:fig1}(c). 

To gather statistics on the dot conductance at each QPC setting, we measure such a conductance trace at 196 different values of $V_{\rm{S1}}$ and $V_{\rm{S2}}$. These are indicated by the black dots superimposed on the conductance map in Fig.~\ref{fig:fig1}(d).  We pick these voltage values such that the sets of trajectories for each shape are independent and such that the area of the dot remains constant at $2.6~\mu\mbox{m}^2$ to within $\pm 5\%$ \cite{SI}. We also compensate for the capacitive effect that changing $V_{\rm{S1}}$ and $V_{\rm{S2}}$ have on the QPC openings by applying corresponding changes in the QPC gate voltages  to maintain exactly one open channel in each lead \cite{SI}. 

\begin{figure}
\begin{center}
\includegraphics[width=8.0cm, keepaspectratio=true]{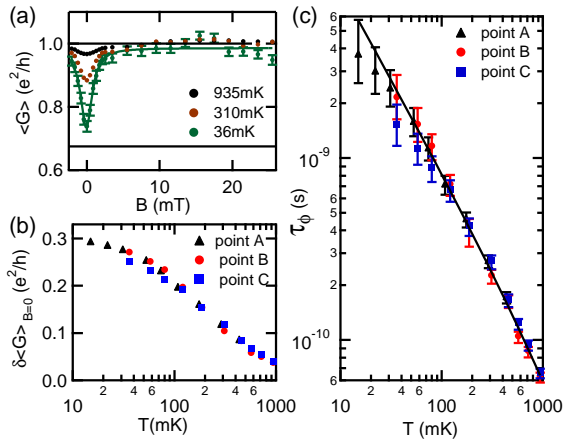}
\end{center}

\caption{(color online) (a) The dot conductance (at QPC setting point C) averaged over 196 different dot shapes as a function of magnetic field at three different temperatures. (b) The average conductance difference between zero and finite magnetic field as function of temperature for the three measured QPC settings indicated in Fig. \ref{fig:fig1}(b). (c) The dephasing time extracted from the weak localization conductance correction as a function of electron temperature for three different QPC settings. The black line is a fit of the combined datasets to the equation discussed in the text.} 
\label{fig:fig2}
\end{figure}

The conductance as a function of magnetic field, averaged over the ensemble of 196 dot shapes, is shown in Fig.~\ref{fig:fig2}(a). The dip in the average conductance at zero magnetic field $\delta \langle G\rangle _{B=0}$ is the signature of coherent backscattering and is referred to as the WL correction.  By fitting a Lorentzian to the magnetic field dependence at each temperature, we extract $\delta \langle G\rangle _{B=0}$ as a function of temperature.  This is shown in Fig.~\ref{fig:fig2}(b) for the three chosen QPC settings.  
Within experimental error (the error bars are of the size of the data markers),  the WL correction of the conductance is insensitive to the precise tuning of the QPCs. 

The analytic expression that connects the size of the WL correction of the average conductance to the dephasing time  is given in \cite{Huibers1999:Thesis} and can be approximated by
\begin{equation}
\delta \langle G\rangle _{B=0}=\frac{1}{2N+1+\gamma} e^2/h
\label{eq:deltag}
\end{equation}
where N=1 is the number of open spin-degenerate modes in each lead and $\gamma=\frac{2\pi\hbar}{\Delta\tau_{\phi}}$ is the dimensionless dephasing rate. The dephasing times extracted using  Eq.~\ref{eq:deltag} are shown in Fig.~\ref{fig:fig2}(c).

For large dephasing, $\gamma \gg 1$, the presence of a small reflection coefficient does not significantly affect the accuracy of its extraction from Eq.~\ref{eq:deltag}, but for dephasing rates of the order of the escape rate $\gamma \approx 2$,  Eq. (\ref{eq:deltag}) cannot account for the effects of a non-zero reflection coefficient \cite{Alves2002, SI}. We therefore also extract the dephasing times using a more complete equation that includes the effects of a reflection coefficient in the QPCs to first order \cite{SI}, and find that the results agree within the experimental error bars with the values in Fig.~\ref{fig:fig2}(c), indicating that the small nonzero reflection coefficient of the QPCs does not introduce any artifacts in this method of determining the dephasing times.

Previous experimental work \cite{Clarke1995, Bird1995,Huibers1999,Micolich2001, Hackens2005} reported the striking observation that the dephasing time did not follow a power law down to the lowest experimentally accessible temperatures and instead appeared to saturate. Fig.~\ref{fig:fig2} shows that when extending the measurements to lower temperatures, in a quantum dot with higher mobility and carefully determined QPC reflection coefficients, the dephasing time based on WL measurements of conductance does not saturate even when $k_B T<\Delta$ or when $\tau_{\phi}>\tau_{\rm{D}}$. This continued increase of the dephasing time with decreasing temperature qualitatively matches theoretical expectations for dephasing caused by electron-electron interactions~\cite{Altshuler1982}.

However, closer examination of the temperature dependence of dephasing time reveals a surprising feature: the dominant contribution is a $T^{-1}$ power law, which does not correspond to the theoretical expectation for
electron-electron interactions in a 0D system (though it does match the behavior seen in other experiments on quantum dots at temperatures above the reported saturation). For a disordered 0D system, explicit calculations predict a power law proportional to $T^{-2}$ ~\cite{Sivan1994, Treiber2009}. In the ballistic 0D case, where large energy exchange processes should dominate, Fermi liquid theory ~\cite{Pines66} is expected to describe the system and the dephasing time should again follow a $T^{-2}$ power law.

As noted earlier, dephasing time power law exponent equal or close to
$-1$ is not only observed in the present experiments but is a common
feature of all measurements of the dephasing time as a function of
temperature in quantum dots ~\cite{Clarke1995, Bird1995,Huibers1998,Huibers1999, Hackens2005}. In some of these quantum
dots ~\cite{Clarke1995,Huibers1999, Hackens2005}, the electron mobility
was low enough that the size of the dot was on the order of or larger than
than the mean free path, so this power law could be attributed to 2D
diffusive behavior. In this case small energy exchange processes yield
a predicted dephasing time inversely proportional to temperature~\cite{Altshuler1982}
\begin{equation}
\tau_{\phi}= \left(\frac{k_B T}{2\pi\hbar} \frac{\lambda_{\rm{F}}}{l_{\rm{elastic}}}\ln \frac{\pi l_{\rm{elastic}}}{\lambda_{\rm{F}}}\right)^{-1}
\label{linearTlaw}
\end{equation}
where $\lambda_{\rm{F}}$ is the electron Fermi wavelength and $l_{\rm elastic}$ is the electron mean free path. Previous experiments did not investigate the value of the exponent in detail.

 In our quantum dot where the  2DEG electron mobility is larger than that in all previous experiments on dephasing in dots, the observation of a $T^{-1}$ power law is striking. To be able to compare directly to the results of previous work we
follow ref.~\cite{Huibers1998} and fit an inverse sum of power laws $(aT+bT^{2})^{-1}$. This fit gives $a=1.2\times 10^{10} $ s$^{-1}$K$^{-1}$ and $b=5.0\times 10^{9}$ s$^{-1}$K$^{-2}$. Previous workers interpreted such fits as
accounting for ballistic and diffusive 2D behavior. In that framework, 
the only free parameter is the mean free path $l_{\rm elastic}$, which can be extracted by fitting the exact form $(aT+bT^{2}$ln$(c/T))^{-1}$. The mean free path, together
with the 2DEG density, determines the coefficient $a$ of the linear-in-T
dephasing term. Coefficients $b$ and $c$ of the quadratic term are both fixed by the
2DEG density. This fit to the
combined data sets for our three QPC settings is shown as the black
trace in Fig.~\ref{fig:fig2}(c). The extracted mean free path $l_{\rm elastic}= 280\pm 11$ nm is close to the value
$l_{\rm elastic}= 250$ nm inferred from previous experiments~\cite{Huibers1998, Huibers19982} on
smaller and larger GaAs quantum dots ($0.4 - 4~\mu \mbox{m}^2$) with substantially lower
bulk 2D mean free path than ours.

The close match of this supposed extracted mean free path between
systems with widely varying dot sizes and bulk mean free paths
suggests that $l_{\rm elastic}$, 4 to 6 times smaller than our dot size,
and 50 times smaller than our bulk 2D mean free path, does not
represent a physical mean free path, but is only a parametrization of
the strength of T-linear dephasing. Indeed, though the combination of
T-linear and quadratic power laws nicely fits measured dephasing over
our entire temperature range, the model that motivated Eq.~\ref{linearTlaw} appears inapplicable to our system, as our quantum dot should be 0D, not 2D,
and is ballistic, not on the border between ballistic and diffusive.

The persistence of a power law behavior of  dephasing rates at low temperatures confirms the theoretical expectation that  dephasing time should diverge as $T\rightarrow 0$, however the value of the power law exponent still needs to be understood. It would be important to determine whether this exponent value is due to a different dephasing mechanism that can also lead to a  $T^{-1}$ law or if the crossover to diffusive 2D behavior occurs at far lower temperature and lower disorder than expected for a given dot size.  An important clue may be that the T-linear contribution to dephasing remains constant across dots with dramatically different sizes and mean free paths~\cite{Huibers1998, Huibers1999}.

\begin{figure}
\begin{center}
\includegraphics[width=8.0cm, keepaspectratio=true]{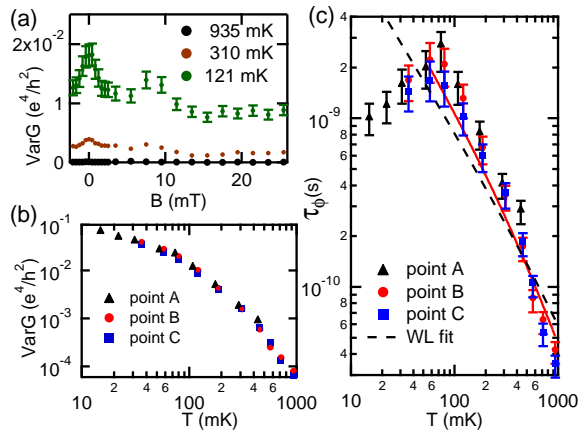}
\end{center}

\caption{(color online) (a) The variance of the conductance (point C) of the ensemble of 196 different dot shapes as a function of magnetic field for three different temperatures. (b) The average conductance variance at finite magnetic field as a function of temperature at the three QPC settings. (c) The dephasing time as a function of electron temperature extracted from the average conductance variance for the three different QPC settings. The red line is a fit  to the sum of power laws of the data for point B from $1$~K to $60$~mK($=T_{CB}$) and the dashed black line is the power law for $\tau_{\phi}$ extracted from the WL data fit (Fig. \ref{fig:fig3}(c)).
}
\label{fig:fig3}
\end{figure}

The temperature-dependence of the dephasing time data extracted from WL is the main result of this work. Dephasing times in quantum dots have also been extracted using the conductance variance Var(G) of the ensemble at various magnetic fields ~\cite{Huibers19982,Hackens2005}. We can perform a similar extraction from our conductance data, and compare to the results of our WL analysis. The magnetic field dependence of Var$G$ is illustrated in Fig.~\ref{fig:fig3}(a) at three different temperatures. To increase statistics we combine the data at several magnetic field values in the range $10-25$ mT for which the dot conductance is uncorrelated and extract the variance of the total set of 784 conductance values  (Fig.~\ref{fig:fig3}(b)). The associated error bars are of the size of the data markers and the good agreement for the three different QPC settings indicates that, similar to the average conductance, the value of Var$G$ is robust against small changes in reflection coefficient.

The variance is affected by temperature in two ways: through the implicit temperature dependence of $\tau_\phi$ and explicitly by thermal averaging. Var$G$ is given by an integral over the conduction correlation function \cite{frahm1995, Hackens2005}.  In the regime where the the total level-broadening (including the effects of dephasing) $\Delta(1+\gamma /2) < 0.6k_BT$,  the temperature effects can be separated from the $\tau_{\phi}$ dependence, and this integral is well-approximated by the expression \cite{Huibers19982}: 
\begin{equation}
\mbox{Var}G_{B \neq 0} =\frac{\Delta(1+\gamma /2)}{6k_B T}(\sqrt{3}+\gamma)^{-2},
\label{eq:varG}
\end{equation}
allowing easy extraction of $\tau_{\phi}$ from Var$G$.
This approximation is applicable in our measurement regime because $\Delta(1+\gamma /2) < 0.6k_BT$ for all but the two lowest temperature points. Fig.~\ref{fig:fig3}(c) shows the dephasing times determined using the data in Fig.~\ref{fig:fig3}(b). We also extract $\tau_{\phi}$ by performing the numerical integration and see that they agree to within the error bars with those extracted using Eq. (\ref{eq:varG}) \cite{SI}.

Let us compare the temperature dependence of the dephasing times extracted from Var$G$ to that extracted from the average conductance (dashed black line in Fig.~\ref{fig:fig3}(c)) Unlike $\tau_{\phi}$ from WL, $\tau_{\phi}$ extracted from the variance abruptly changes at $T\approx 60$ mK. At temperatures larger than $60$ mK, the two methods yield a similar power law  and  the fit to $\tau_{\phi}$ based on Var$G$,  shown as a red trace in fig.~\ref{fig:fig3}(c)), yields a mean free path $l_{\rm elastic}=500\pm 70$ nm. At lower temperatures we observe a marked decrease of $\tau_{\phi}$ inferred from Var$G$, the onset of which coincides with the temperature at which mesoscopic Coulomb Blockade (MCB) oscillations in conductance appear superimposed on the UCFs \cite{Sami}. MCB should affect the variance of the conductance, but counterintuitively is predicted not to affect the average conductance \cite{Brouwer2005:NonequilCBThy, Brouwer2005:WeakCBinDots}. In the absence of a theoretical analysis of this regime, we are unable to separate the contributions of the CB oscillations and those of the UCFs to the ensemble variance. As a result, we conclude that this second method for determining $\tau_{\phi}$ is less reliable at low temperatures than the WL method.

In conclusion, measurements of the average conductance of an ensemble of shapes of a quantum dot tuned to near-perfect QPC transmission, can be used to extract an electron dephasing time that does not show saturation down to temperatures below those of previous experiments. This experimental observation establishes that the dephasing time in a quantum dot continues to increase with decreasing temperature as theoretically predicted.

\begin{acknowledgements}
	We are grateful to P. W. Brouwer, J. von Delft, Y. Oreg, I. L. Aleiner , C. Lewenkopf, D. Cohen  and Y. Imry for discussions. This work was supported by the NSF under DMR-0906062 and  the U.S.-Israel BSF grant No. 2008149.
\end{acknowledgements}
	

\begin{thebibliography}{34}%
\makeatletter
\providecommand \@ifxundefined [1]{%
 \ifx #1\undefined \expandafter \@firstoftwo
 \else \expandafter \@secondoftwo
\fi
}%
\providecommand \@ifnum [1]{%
 \ifnum #1\expandafter \@firstoftwo
 \else \expandafter \@secondoftwo
\fi
}%
\providecommand \enquote [1]{``#1''}%
\providecommand \bibnamefont  [1]{#1}%
\providecommand \bibfnamefont [1]{#1}%
\providecommand \citenamefont [1]{#1}%
\providecommand\href[0]{\@sanitize\@href}%
\providecommand\@href[1]{\endgroup\@@startlink{#1}\endgroup\@@href}%
\providecommand\@@href[1]{#1\@@endlink}%
\providecommand \@sanitize [0]{\begingroup\catcode`\&12\catcode`\#12\relax}%
\@ifxundefined \pdfoutput {\@firstoftwo}{%
 \@ifnum{\z@=\pdfoutput}{\@firstoftwo}{\@secondoftwo}%
}{%
 \providecommand\@@startlink[1]{\leavevmode}%
 \providecommand\@@endlink[0]{}%
}{%
 \providecommand\@@startlink[1]{%
  \leavevmode
  \pdfstartlink
   attr{/Border[0 0 1 ]/H/I/C[0 1 1]}%
   user{/Subtype/Link/A<</Type/Action/S/URI/URI(#1)>>}%
  \relax
 }%
 \providecommand\@@endlink[0]{\pdfendlink}%
}%
\providecommand \url  [0]{\begingroup\@sanitize \@url }%
\providecommand \@url [1]{\endgroup\@href {#1}{\urlprefix}}%
\providecommand \urlprefix [0]{URL }%
\providecommand \Eprint[0]{\href }%
\@ifxundefined \urlstyle {%
  \providecommand \doi [1]{doi:\discretionary{}{}{}#1}%
}{%
  \providecommand \doi [0]{doi:\discretionary{}{}{}\begingroup
  \urlstyle{rm}\Url }%
}%
\providecommand \doibase [0]{http://dx.doi.org/}%
\providecommand \Doi[1]{\href{\doibase#1}}%
\providecommand \bibAnnote [3]{%
  \BibitemShut{#1}%
  \begin{quotation}\noindent
    \textsc{Key:}\ #2\\\textsc{Annotation:}\ #3%
  \end{quotation}%
}%
\providecommand \bibAnnoteFile [2]{%
  \IfFileExists{#2}{\bibAnnote {#1} {#2} {\input{#2}}}{}%
}%
\providecommand \typeout [0]{\immediate \write \m@ne }%
\providecommand \selectlanguage [0]{\@gobble}%
\providecommand \bibinfo [0]{\@secondoftwo}%
\providecommand \bibfield [0]{\@secondoftwo}%
\providecommand \translation [1]{[#1]}%
\providecommand \BibitemOpen[0]{}%
\providecommand \bibitemStop [0]{}%
\providecommand \bibitemNoStop [0]{.\EOS\space}%
\providecommand \EOS [0]{\spacefactor3000\relax}%
\providecommand \BibitemShut [1]{\csname bibitem#1\endcsname}%
\bibitem{Altshuler1985}%
  \BibitemOpen
  \bibfield{author}{%
  \bibinfo {author} {\bibfnamefont{B.~L.}\ \bibnamefont{Altshuler}}\ and\
  \bibinfo {author} {\bibfnamefont{A.~G.}\ \bibnamefont{Aronov}},\ }%
  \enquote{\bibinfo {title} {Electron--electron interaction in disordered
  conductors},}\ \ (\bibinfo {publisher} {North--Holland},\ \bibinfo {address}
  {Amsterdam},\ \bibinfo {year} {1985})\ Chap.~\bibinfo {chapter} {1}, pp.\
  \bibinfo {pages} {1 -- 153}%
  \bibAnnoteFile{NoStop}{Altshuler1985}%
\bibitem{Zheng1996}%
  \BibitemOpen
  \bibfield{author}{%
  \bibinfo {author} {\bibfnamefont{L.}~\bibnamefont{Zheng}}\ and\ \bibinfo
  {author} {\bibfnamefont{S.}~\bibnamefont{Das~Sarma}},\ }%
  \bibfield{journal}{%
  \Doi{10.1103/PhysRevB.53.9964}{\bibinfo {journal} {Phys. Rev. B}}\ }%
  \textbf{\bibinfo {volume} {53}},\ \bibinfo {pages} {9964} (\bibinfo {year}
  {1996})%
  \bibAnnoteFile{NoStop}{Zheng1996}%
\bibitem{Altshuler1982}%
  \BibitemOpen
  \bibfield{author}{%
  \bibinfo {author} {\bibfnamefont{B.~L.}\ \bibnamefont{Altshuler}}, \bibinfo
  {author} {\bibfnamefont{A.~G.}\ \bibnamefont{Aronov}},\ and\ \bibinfo
  {author} {\bibfnamefont{D.~E.}\ \bibnamefont{Khmelnitskii}},\ }%
  \bibfield{journal}{%
  \bibinfo {journal} {J. Phys. C: Solid State Phys.}\ }%
  \textbf{\bibinfo {volume} {15}},\ \bibinfo {pages} {7367} (\bibinfo {year}
  {1982})%
  \bibAnnoteFile{NoStop}{Altshuler1982}%
\bibitem{Stern1990}%
  \BibitemOpen
  \bibfield{author}{%
  \bibinfo {author} {\bibfnamefont{A.}~\bibnamefont{Stern}}, \bibinfo {author}
  {\bibfnamefont{Y.}~\bibnamefont{Aharonov}},\ and\ \bibinfo {author}
  {\bibfnamefont{Y.}~\bibnamefont{Imry}},\ }%
  \bibfield{journal}{%
  \Doi{10.1103/PhysRevA.41.3436}{\bibinfo {journal} {Phys. Rev. A}}\ }%
  \textbf{\bibinfo {volume} {41}},\ \bibinfo {pages} {3436} (\bibinfo {year}
  {1990})%
  \bibAnnoteFile{NoStop}{Stern1990}%
\bibitem{Clarke1995}%
  \BibitemOpen
  \bibfield{author}{%
  \bibinfo {author} {\bibfnamefont{R.~M.}\ \bibnamefont{Clarke}}, \bibinfo
  {author} {\bibfnamefont{I.~H.}\ \bibnamefont{Chan}}, \bibinfo {author}
  {\bibfnamefont{C.~M.}\ \bibnamefont{Marcus}}, \bibinfo {author}
  {\bibfnamefont{C.~I.}\ \bibnamefont{Duru\"oz}}, \bibinfo {author}
  {\bibfnamefont{J.~S.}\ \bibnamefont{Harris}}, \bibinfo {author}
  {\bibfnamefont{K.}~\bibnamefont{Campman}},\ and\ \bibinfo {author}
  {\bibfnamefont{A.~C.}\ \bibnamefont{Gossard}},\ }%
  \bibfield{journal}{%
  \bibinfo {journal} {Phys. Rev. B}\ }%
  \textbf{\bibinfo {volume} {52}},\ \bibinfo {pages} {2656} (\bibinfo {year}
  {1995})%
  \bibAnnoteFile{NoStop}{Clarke1995}%
\bibitem{Bird1995}%
  \BibitemOpen
  \bibfield{author}{%
  \bibinfo {author} {\bibfnamefont{J.~P.}\ \bibnamefont{Bird}}, \bibinfo
  {author} {\bibfnamefont{K.}~\bibnamefont{Ishibashi}}, \bibinfo {author}
  {\bibfnamefont{D.~K.}\ \bibnamefont{Ferry}}, \bibinfo {author}
  {\bibfnamefont{Y.}~\bibnamefont{Ochiai}}, \bibinfo {author}
  {\bibfnamefont{Y.}~\bibnamefont{Aoyagi}},\ and\ \bibinfo {author}
  {\bibfnamefont{T.}~\bibnamefont{Sugano}},\ }%
  \bibfield{journal}{%
  \bibinfo {journal} {Phys. Rev. B}\ }%
  \textbf{\bibinfo {volume} {51}},\ \bibinfo {pages} {18037} (\bibinfo {year}
  {1995})%
  \bibAnnoteFile{NoStop}{Bird1995}%
\bibitem{Huibers1999}%
  \BibitemOpen
  \bibfield{author}{%
  \bibinfo {author} {\bibfnamefont{A.~G.}\ \bibnamefont{Huibers}}, \bibinfo
  {author} {\bibfnamefont{J.~A.}\ \bibnamefont{Folk}}, \bibinfo {author}
  {\bibfnamefont{S.~R.}\ \bibnamefont{Patel}}, \bibinfo {author}
  {\bibfnamefont{C.~M.}\ \bibnamefont{Marcus}}, \bibinfo {author}
  {\bibfnamefont{C.~I.}\ \bibnamefont{Duru\"oz}},\ and\ \bibinfo {author}
  {\bibfnamefont{J.~S.}\ \bibnamefont{Harris}},\ }%
  \bibfield{journal}{%
  \bibinfo {journal} {Phys. Rev. Lett.}\ }%
  \textbf{\bibinfo {volume} {83}},\ \bibinfo {pages} {5090} (\bibinfo {year}
  {1999})%
  \bibAnnoteFile{NoStop}{Huibers1999}%
\bibitem{Micolich2001}%
  \BibitemOpen
  \bibfield{author}{%
  \bibinfo {author} {\bibfnamefont{A.~P.}\ \bibnamefont{Micolich}}, \bibinfo
  {author} {\bibfnamefont{R.~P.}\ \bibnamefont{Taylor}}, \bibinfo {author}
  {\bibfnamefont{A.~G.}\ \bibnamefont{Davies}}, \bibinfo {author}
  {\bibfnamefont{J.~P.}\ \bibnamefont{Bird}}, \bibinfo {author}
  {\bibfnamefont{R.}~\bibnamefont{Newbury}}, \bibinfo {author}
  {\bibfnamefont{T.~M.}\ \bibnamefont{Fromhold}}, \bibinfo {author}
  {\bibfnamefont{A.}~\bibnamefont{Ehlert}}, \bibinfo {author}
  {\bibfnamefont{H.}~\bibnamefont{Linke}}, \bibinfo {author}
  {\bibfnamefont{L.~D.}\ \bibnamefont{Macks}}, \bibinfo {author}
  {\bibfnamefont{W.~R.}\ \bibnamefont{Tribe}}, \bibinfo {author}
  {\bibfnamefont{E.~H.}\ \bibnamefont{Linfield}}, \bibinfo {author}
  {\bibfnamefont{D.~A.}\ \bibnamefont{Ritchie}}, \bibinfo {author}
  {\bibfnamefont{J.}~\bibnamefont{Cooper}}, \bibinfo {author}
  {\bibfnamefont{Y.}~\bibnamefont{Aoyagi}},\ and\ \bibinfo {author}
  {\bibfnamefont{P.~B.}\ \bibnamefont{Wilkinson}},\ }%
  \bibfield{journal}{%
  \bibinfo {journal} {Phys. Rev. Lett.}\ }%
  \textbf{\bibinfo {volume} {87}},\ \bibinfo {pages} {036802} (\bibinfo {year}
  {2001})%
  \bibAnnoteFile{NoStop}{Micolich2001}%
\bibitem{Hackens2005}%
  \BibitemOpen
  \bibfield{author}{%
  \bibinfo {author} {\bibfnamefont{B.}~\bibnamefont{Hackens}}, \bibinfo
  {author} {\bibfnamefont{S.}~\bibnamefont{Faniel}}, \bibinfo {author}
  {\bibfnamefont{C.}~\bibnamefont{Gustin}}, \bibinfo {author}
  {\bibfnamefont{X.}~\bibnamefont{Wallart}}, \bibinfo {author}
  {\bibfnamefont{S.}~\bibnamefont{Bollaert}}, \bibinfo {author}
  {\bibfnamefont{A.}~\bibnamefont{Cappy}},\ and\ \bibinfo {author}
  {\bibfnamefont{V.}~\bibnamefont{Bayot}},\ }%
  \bibfield{journal}{%
  \bibinfo {journal} {Phys. Rev. Lett.}\ }%
  \textbf{\bibinfo {volume} {94}},\ \bibinfo {pages} {146802} (\bibinfo {year}
  {2005})%
  \bibAnnoteFile{NoStop}{Hackens2005}%
\bibitem{Eshkol2006}%
  \BibitemOpen
  \bibfield{author}{%
  \bibinfo {author} {\bibfnamefont{M.}~\bibnamefont{Eshkol}}, \bibinfo {author}
  {\bibfnamefont{E.}~\bibnamefont{Eisenberg}}, \bibinfo {author}
  {\bibfnamefont{M.}~\bibnamefont{Karpovski}},\ and\ \bibinfo {author}
  {\bibfnamefont{A.}~\bibnamefont{Palevski}},\ }%
  \bibfield{journal}{%
  \bibinfo {journal} {Phys. Rev. B}\ }%
  \textbf{\bibinfo {volume} {73}},\ \bibinfo {pages} {115318} (\bibinfo {year}
  {2006})%
  \bibAnnoteFile{NoStop}{Eshkol2006}%
\bibitem{Mohanty1997}%
  \BibitemOpen
  \bibfield{author}{%
  \bibinfo {author} {\bibfnamefont{P.}~\bibnamefont{Mohanty}}, \bibinfo
  {author} {\bibfnamefont{E.~M.~Q.}\ \bibnamefont{Jariwala}},\ and\ \bibinfo
  {author} {\bibfnamefont{R.~A.}\ \bibnamefont{Webb}},\ }%
  \bibfield{journal}{%
  \bibinfo {journal} {Phys. Rev. Lett.}\ }%
  \textbf{\bibinfo {volume} {78}},\ \bibinfo {pages} {3366} (\bibinfo {year}
  {1997})%
  \bibAnnoteFile{NoStop}{Mohanty1997}%
\bibitem{Anthore2003}%
  \BibitemOpen
  \bibfield{author}{%
  \bibinfo {author} {\bibfnamefont{A.}~\bibnamefont{Anthore}}, \bibinfo
  {author} {\bibfnamefont{F.}~\bibnamefont{Pierre}}, \bibinfo {author}
  {\bibfnamefont{H.}~\bibnamefont{Pothier}},\ and\ \bibinfo {author}
  {\bibfnamefont{D.}~\bibnamefont{Esteve}},\ }%
  \bibfield{journal}{%
  \bibinfo {journal} {Phys. Rev. Lett.}\ }%
  \textbf{\bibinfo {volume} {90}},\ \bibinfo {pages} {076806} (\bibinfo {year}
  {2003})%
  \bibAnnoteFile{NoStop}{Anthore2003}%
\bibitem{Pierre2003}%
  \BibitemOpen
  \bibfield{author}{%
  \bibinfo {author} {\bibfnamefont{F.}~\bibnamefont{Pierre}}, \bibinfo {author}
  {\bibfnamefont{A.~B.}\ \bibnamefont{Gougam}}, \bibinfo {author}
  {\bibfnamefont{A.}~\bibnamefont{Anthore}}, \bibinfo {author}
  {\bibfnamefont{H.}~\bibnamefont{Pothier}}, \bibinfo {author}
  {\bibfnamefont{D.}~\bibnamefont{Esteve}},\ and\ \bibinfo {author}
  {\bibfnamefont{N.~O.}\ \bibnamefont{Birge}},\ }%
  \bibfield{journal}{%
  \bibinfo {journal} {Phys. Rev. B}\ }%
  \textbf{\bibinfo {volume} {68}},\ \bibinfo {pages} {085413} (\bibinfo {year}
  {2003})%
  \bibAnnoteFile{NoStop}{Pierre2003}%
\bibitem{Narozhny2002}%
  \BibitemOpen
  \bibfield{author}{%
  \bibinfo {author} {\bibfnamefont{B.~N.}\ \bibnamefont{Narozhny}}, \bibinfo
  {author} {\bibfnamefont{G.}~\bibnamefont{Zala}},\ and\ \bibinfo {author}
  {\bibfnamefont{I.~L.}\ \bibnamefont{Aleiner}},\ }%
  \bibfield{journal}{%
  \Doi{10.1103/PhysRevB.65.180202}{\bibinfo {journal} {Phys. Rev. B}}\ }%
  \textbf{\bibinfo {volume} {65}},\ \bibinfo {pages} {180202} (\bibinfo {year}
  {2002})%
  \bibAnnoteFile{NoStop}{Narozhny2002}%
\bibitem{Golubev1998}%
  \BibitemOpen
  \bibfield{author}{%
  \bibinfo {author} {\bibfnamefont{D.~S.}\ \bibnamefont{Golubev}}\ and\
  \bibinfo {author} {\bibfnamefont{A.~D.}\ \bibnamefont{Zaikin}},\ }%
  \bibfield{journal}{%
  \Doi{10.1103/PhysRevLett.81.1074}{\bibinfo {journal} {Phys. Rev. Lett.}}\ }%
  \textbf{\bibinfo {volume} {81}},\ \bibinfo {pages} {1074} (\bibinfo {year}
  {1998})%
  \bibAnnoteFile{NoStop}{Golubev1998}%
\bibitem{Cedraschi2000}%
  \BibitemOpen
  \bibfield{author}{%
  \bibinfo {author} {\bibfnamefont{P.}~\bibnamefont{Cedraschi}}, \bibinfo
  {author} {\bibfnamefont{V.~V.}\ \bibnamefont{Ponomarenko}},\ and\ \bibinfo
  {author} {\bibfnamefont{M.}~\bibnamefont{B\"uttiker}},\ }%
  \bibfield{journal}{%
  \Doi{10.1103/PhysRevLett.84.346}{\bibinfo {journal} {Phys. Rev. Lett.}}\ }%
  \textbf{\bibinfo {volume} {84}},\ \bibinfo {pages} {346} (\bibinfo {year}
  {2000})%
  \bibAnnoteFile{NoStop}{Cedraschi2000}%
\bibitem{Aleiner1998}%
  \BibitemOpen
  \bibfield{author}{%
  \bibinfo {author} {\bibfnamefont{B.~L.}\ \bibnamefont{Altshuler}}, \bibinfo
  {author} {\bibfnamefont{M.~E.}\ \bibnamefont{Gershenson}},\ and\ \bibinfo
  {author} {\bibfnamefont{I.~L.}\ \bibnamefont{Aleiner}},\ }%
  \bibfield{journal}{%
  \bibinfo {journal} {Physica (Amsterdam)}\ }%
  \textbf{\bibinfo {volume} {3E}},\ \bibinfo {pages} {58} (\bibinfo {year}
  {1998})%
  \bibAnnoteFile{NoStop}{Aleiner1998}%
\bibitem{Aleiner1999}%
  \BibitemOpen
  \bibfield{author}{%
  \bibinfo {author} {\bibfnamefont{I.~L.}\ \bibnamefont{Aleiner}}, \bibinfo
  {author} {\bibfnamefont{B.~L.}\ \bibnamefont{Altshuler}},\ and\ \bibinfo
  {author} {\bibfnamefont{M.~E.}\ \bibnamefont{Gershenson}},\ }%
  \bibfield{journal}{%
  \bibinfo {journal} {Phys. Rev. Lett.}\ }%
  \textbf{\bibinfo {volume} {82}},\ \bibinfo {pages} {3190} (\bibinfo {year}
  {1999})%
  \bibAnnoteFile{NoStop}{Aleiner1999}%
\bibitem{Goppert2002}%
  \BibitemOpen
  \bibfield{author}{%
  \bibinfo {author} {\bibfnamefont{G.}~\bibnamefont{G\"oppert}}, \bibinfo
  {author} {\bibfnamefont{Y.~M.}\ \bibnamefont{Galperin}}, \bibinfo {author}
  {\bibfnamefont{B.~L.}\ \bibnamefont{Altshuler}},\ and\ \bibinfo {author}
  {\bibfnamefont{H.}~\bibnamefont{Grabert}},\ }%
  \bibfield{journal}{%
  \Doi{10.1103/PhysRevB.66.195328}{\bibinfo {journal} {Phys. Rev. B}}\ }%
  \textbf{\bibinfo {volume} {66}},\ \bibinfo {pages} {195328} (\bibinfo {year}
  {2002})%
  \bibAnnoteFile{NoStop}{Goppert2002}%
\bibitem{Zarand2004}%
  \BibitemOpen
  \bibfield{author}{%
  \bibinfo {author} {\bibfnamefont{G.}~\bibnamefont{Zar\'and}}, \bibinfo
  {author} {\bibfnamefont{L.}~\bibnamefont{Borda}}, \bibinfo {author}
  {\bibfnamefont{J.}~\bibnamefont{von Delft}},\ and\ \bibinfo {author}
  {\bibfnamefont{N.}~\bibnamefont{Andrei}},\ }%
  \bibfield{journal}{%
  \Doi{10.1103/PhysRevLett.93.107204}{\bibinfo {journal} {Phys. Rev. Lett.}}\
  }%
  \textbf{\bibinfo {volume} {93}},\ \bibinfo {pages} {107204} (\bibinfo {year}
  {2004})%
  \bibAnnoteFile{NoStop}{Zarand2004}%
\bibitem{Huibers1998}%
  \BibitemOpen
  \bibfield{author}{%
  \bibinfo {author} {\bibfnamefont{A.~G.}\ \bibnamefont{Huibers}}, \bibinfo
  {author} {\bibfnamefont{M.}~\bibnamefont{Switkes}}, \bibinfo {author}
  {\bibfnamefont{C.~M.}\ \bibnamefont{Marcus}}, \bibinfo {author}
  {\bibfnamefont{K.}~\bibnamefont{Campman}},\ and\ \bibinfo {author}
  {\bibfnamefont{A.~C.}\ \bibnamefont{Gossard}},\ }%
  \bibfield{journal}{%
  \Doi{10.1103/PhysRevLett.81.200}{\bibinfo {journal} {Phys. Rev. Lett.}}\ }%
  \textbf{\bibinfo {volume} {81}},\ \bibinfo {pages} {200} (\bibinfo {year}
  {1998})%
  \bibAnnoteFile{NoStop}{Huibers1998}%
\bibitem{Huibers19982}%
  \BibitemOpen
  \bibfield{author}{%
  \bibinfo {author} {\bibfnamefont{A.~G.}\ \bibnamefont{Huibers}}, \bibinfo
  {author} {\bibfnamefont{S.~R.}\ \bibnamefont{Patel}}, \bibinfo {author}
  {\bibfnamefont{C.~M.}\ \bibnamefont{Marcus}}, \bibinfo {author}
  {\bibfnamefont{P.~W.}\ \bibnamefont{Brouwer}}, \bibinfo {author}
  {\bibfnamefont{C.~I.}\ \bibnamefont{Duru\"oz}},\ and\ \bibinfo {author}
  {\bibfnamefont{J.~S.}\ \bibnamefont{Harris}},\ }%
  \bibfield{journal}{%
  \bibinfo {journal} {Phys. Rev. Lett.}\ }%
  \textbf{\bibinfo {volume} {81}},\ \bibinfo {pages} {1917} (\bibinfo {year}
  {1998})%
  \bibAnnoteFile{NoStop}{Huibers19982}%
\bibitem{beenakker.vanhouten}%
  \BibitemOpen
  \bibfield{author}{%
  \bibinfo {author} {\bibfnamefont{C.}~\bibnamefont{Beenakker}}\ and\ \bibinfo
  {author} {\bibfnamefont{H.}~\bibnamefont{van Houten}},\ }%
  in\ \emph{\bibinfo {booktitle} {Solid State Physics}},\ Vol.~\bibinfo
  {volume} {44},\ \bibinfo {editor} {edited by\ \bibinfo {editor}
  {\bibfnamefont{H.}~\bibnamefont{Ehrenreich}}\ and\ \bibinfo {editor}
  {\bibfnamefont{D.}~\bibnamefont{Turnbull}}}\ (\bibinfo {publisher} {Academic
  Press},\ \bibinfo {year} {1991})%
  \bibAnnoteFile{NoStop}{beenakker.vanhouten}%
\bibitem{Brouwer1997}%
  \BibitemOpen
  \bibfield{author}{%
  \bibinfo {author} {\bibfnamefont{P.~W.}\ \bibnamefont{Brouwer}}\ and\
  \bibinfo {author} {\bibfnamefont{C.~W.~J.}\ \bibnamefont{Beenakker}},\ }%
  \bibfield{journal}{%
  \Doi{10.1103/PhysRevB.55.4695}{\bibinfo {journal} {Phys. Rev. B}}\ }%
  \textbf{\bibinfo {volume} {55}},\ \bibinfo {pages} {4695} (\bibinfo {year}
  {1997})%
  \bibAnnoteFile{NoStop}{Brouwer1997}%
\bibitem{SI}%
  \BibitemOpen
  \bibinfo {note} {Please see supplementary material in EPAPS Document No. at
  http}%
  \bibAnnoteFile{NoStop}{SI}%
\bibitem{Huibers1999:Thesis}%
  \BibitemOpen
  \bibfield{author}{%
  \bibinfo {author} {\bibfnamefont{A.~G.}\ \bibnamefont{Huibers}},\ }%
  \bibinfo {note} {\protect{P}h.D. Thesis, Stanford University (1999)}%
  \bibAnnoteFile{NoStop}{Huibers1999:Thesis}%
\bibitem{Alves2002}%
  \BibitemOpen
  \bibfield{author}{%
  \bibinfo {author} {\bibfnamefont{E.~R.~P.}\ \bibnamefont{Alves}}\ and\
  \bibinfo {author} {\bibfnamefont{C.~H.}\ \bibnamefont{Lewenkopf}},\ }%
  \bibfield{journal}{%
  \Doi{10.1103/PhysRevLett.88.256805}{\bibinfo {journal} {Phys. Rev. Lett.}}\
  }%
  \textbf{\bibinfo {volume} {88}},\ \bibinfo {pages} {256805} (\bibinfo {year}
  {2002})%
  \bibAnnoteFile{NoStop}{Alves2002}%
\bibitem{Sivan1994}%
  \BibitemOpen
  \bibfield{author}{%
  \bibinfo {author} {\bibfnamefont{U.}~\bibnamefont{Sivan}}, \bibinfo {author}
  {\bibfnamefont{Y.}~\bibnamefont{Imry}},\ and\ \bibinfo {author}
  {\bibfnamefont{A.~G.}\ \bibnamefont{Aronov}},\ }%
  \bibfield{journal}{%
  \bibinfo {journal} {Europhys. Lett.}\ }%
  \textbf{\bibinfo {volume} {28}},\ \bibinfo {pages} {115} (\bibinfo {year}
  {1994})%
  \bibAnnoteFile{NoStop}{Sivan1994}%
\bibitem{Treiber2009}%
  \BibitemOpen
  \bibfield{author}{%
  \bibinfo {author} {\bibfnamefont{M.}~\bibnamefont{Treiber}}, \bibinfo
  {author} {\bibfnamefont{O.~M.}\ \bibnamefont{Yevtushenko}}, \bibinfo {author}
  {\bibfnamefont{F.}~\bibnamefont{Marquardt}}, \bibinfo {author}
  {\bibfnamefont{J.}~\bibnamefont{von Delft}},\ and\ \bibinfo {author}
  {\bibfnamefont{I.~V.}\ \bibnamefont{Lerner}},\ }%
  \bibfield{journal}{%
  \Doi{10.1103/PhysRevB.80.201305}{\bibinfo {journal} {Phys. Rev. B}}\ }%
  \textbf{\bibinfo {volume} {80}},\ \bibinfo {pages} {201305} (\bibinfo {year}
  {2009})%
  \bibAnnoteFile{NoStop}{Treiber2009}%
\bibitem{Pines66}%
  \BibitemOpen
  \bibfield{author}{%
  \bibinfo {author} {\bibfnamefont{D.}~\bibnamefont{Pines}}\ and\ \bibinfo
  {author} {\bibfnamefont{P.}~\bibnamefont{Nozi{\`e}res}},\ }%
  \emph{\bibinfo {title} {The theory of quantum liquids}}\ (\bibinfo
  {publisher} {W.~A.~Benjamin Inc},\ \bibinfo {address} {New York},\ \bibinfo
  {year} {1966})%
  \bibAnnoteFile{NoStop}{Pines66}%
\bibitem{frahm1995}%
  \BibitemOpen
  \bibfield{author}{%
  \bibinfo {author} {\bibfnamefont{K.}~\bibnamefont{Frahm}},\ }%
  \bibfield{journal}{%
  \bibinfo {journal} {Europhys. Letters}\ }%
  \textbf{\bibinfo {volume} {30}},\ \bibinfo {pages} {457} (\bibinfo {year}
  {1995})%
  \bibAnnoteFile{NoStop}{frahm1995}%
\bibitem{Sami}%
  \BibitemOpen
  \bibfield{author}{%
  \bibinfo {author} {\bibfnamefont{S.}~\bibnamefont{Amasha}}, \bibinfo {author}
  {\bibfnamefont{I.~G.}\ \bibnamefont{Rau}}, \bibinfo {author}
  {\bibfnamefont{M.}~\bibnamefont{Grobis}}, \bibinfo {author}
  {\bibfnamefont{R.~M.}\ \bibnamefont{Potok}}, \bibinfo {author}
  {\bibfnamefont{H.}~\bibnamefont{Shtrikman}},\ and\ \bibinfo {author}
  {\bibfnamefont{D.}~\bibnamefont{Goldhaber-Gordon}},\ }%
  \bibinfo {journal} {Phys. Rev. Lett.}%
  \bibAnnoteFile{Stop}{Sami}%
\bibitem{Brouwer2005:NonequilCBThy}%
  \BibitemOpen
\bibfield{journal}{%
    }%
  \bibfield{author}{%
  \bibinfo {author} {\bibfnamefont{P.~W.}\ \bibnamefont{Brouwer}}, \bibinfo
  {author} {\bibfnamefont{A.}~\bibnamefont{Lamacraft}},\ and\ \bibinfo {author}
  {\bibfnamefont{K.}~\bibnamefont{Flensberg}},\ }%
  \bibfield{journal}{%
  \bibinfo {journal} {Phys. Rev. B}\ }%
  \textbf{\bibinfo {volume} {72}},\ \bibinfo {pages} {075316} (\bibinfo {year}
  {2005})%
  \bibAnnoteFile{NoStop}{Brouwer2005:NonequilCBThy}%
\bibitem{Brouwer2005:WeakCBinDots}%
  \BibitemOpen
  \bibfield{author}{%
  \bibinfo {author} {\bibfnamefont{P.~W.}\ \bibnamefont{Brouwer}}, \bibinfo
  {author} {\bibfnamefont{A.}~\bibnamefont{Lamacraft}},\ and\ \bibinfo {author}
  {\bibfnamefont{K.}~\bibnamefont{Flensberg}},\ }%
  \bibfield{journal}{%
  \Doi{10.1103/PhysRevLett.94.136801}{\bibinfo {journal} {Phys. Rev. Lett.}}\
  }%
  \textbf{\bibinfo {volume} {94}},\ \bibinfo {pages} {136801} (\bibinfo {year}
  {2005})%
  \bibAnnoteFile{NoStop}{Brouwer2005:WeakCBinDots}%
\end{thebibliography}

\newcommand{\noopsort}[1]{} \newcommand{\printfirst}[2]{#1}
  \newcommand{\singleletter}[1]{#1} \newcommand{\switchargs}[2]{#2#1}

\end{document}